\documentclass[aps,preprintnumbers,10pt,superscriptaddress,twocolumn,nofootinbib]{revtex4-1}
\usepackage[dvips]{graphicx}

\usepackage{siunitx}
\usepackage{array,amsmath,amsthm,feynmf,mathtools}
\usepackage{multirow}
\usepackage{mathtools}
\usepackage{epsfig}
\usepackage{graphicx}
\usepackage{xcolor}
\usepackage{siunitx}
\usepackage{slashed}
\usepackage{amssymb}
\usepackage{bm}
\usepackage[export]{adjustbox}
\usepackage{enumitem}
\usepackage[colorlinks=true,linkcolor=blue,citecolor=blue]{hyperref}

\definecolor{col1}{rgb}{0.858, 0, 0}
\definecolor{col2}{rgb}{0, 0.458, 0}
\definecolor{col3}{rgb}{0, 0, 0.858}

\newcommand{\mtxY}[0]{\mathbb{Y}}
\newcommand{\argdet}[0]{\text{arg\,det}}
\newcommand{\imtr}[0]{\text{Im\,Tr}}

\begin{document}

\title{Do Minimal Parity Solutions to the Strong $CP$ Problem Work?}
\author{Jordy de Vries}
\email{j.devries4@uva.nl}
\affiliation{ Institute for Theoretical Physics Amsterdam and Delta Institute for Theoretical Physics, University of Amsterdam, Science Park 904, 1098 XH Amsterdam, The Netherlands}
\affiliation{Nikhef, Theory Group, Science Park 105, 1098 XG, Amsterdam, The Netherlands}

\author{Patrick Draper}
\email{pdraper@illinois.edu}
\affiliation{Department of Physics, University of Illinois, Urbana, IL 61801}

\author{Hiren H. Patel}
\email{hpatel6@ucsc.edu}
\affiliation{Department of Physics and Santa Cruz Institute for Particle Physics\\ University of California, Santa Cruz, CA 95064, USA}

%%%%%%%%%%%%%%%%%%%%%%%%%%%%%%%%%%%%%%%%%%%%%%%%%%%%%%%%%%%%%%%%%%%%
\begin{abstract}
One class of solutions to the strong $CP$ problem relies on generalized parity symmetries. A minimal model of this type, constructed by Babu and Mohapatra and based on a softly broken parity symmetry, has the remarkable property that effective QCD vacuum angle $\bar\theta$ vanishes up to one-loop order.  
We compute the leading two-loop contributions to $\bar\theta$ in this model and estimate subleading contributions. In contrast to previous estimates, we argue that $\bar \theta$ is not suppressed by the weak scale, and  we find contributions of order $10^{-3}\mbox{--}10^{-2}$ multiplying unknown mixing angles and phases. Thus the model does not generically address the strong $CP$ problem, but it might be made consistent with $\bar\theta<10^{-10}$ in some corners of parameter space. For such non-generic parameters, $\bar\theta$ is still likely to be just below present bounds, and therefore provides the dominant source of hadronic EDMs. We discuss the resulting EDM phenomenology.
\end{abstract}
%%%%%%%%%%%%%%%%%%%%%%%%%%%%%%%%%%%%%%%%%%%%%%%%%%%%%%%%%%%%%%%%%%%%
%\pacs{}

\maketitle
%%%%%%%%%%%%%%%%%%%%%%%%%%%%%%%%%%%%%%%%%%%%%%%%%%%%%%%%%%%%%%%%%%%%%
\section{Introduction}

Generalized parity symmetries provide an interesting mechanism for addressing the strong $CP$ problem~\cite{mohapatrasenjanovic,Beg:1978mt,Georgi:1978xz,Babu:1989rb,barrsenjanovic}. If the  ultraviolet  completion of the Standard Model (SM) respects a suitable parity symmetry, the bare QCD vacuum angle $\bar\theta$ vanishes. In such models, parity must be broken at some lower scale, but still above the weak scale for compatibility with experiment. The trick in these models, as in all solutions to the strong $CP$ problem relying on UV symmetries~\cite{Nelson:1983zb,Barr:1984qx,bbp,hillerschmaltz}, is to engineer this breaking in a way that does not radiatively generate a large $\bar\theta$ term~\cite{Albaid:2015axa}.

The simplest model realizing this idea was constructed by Babu and Mohapatra~\cite{Babu:1989rb}. Here the  $SU(2)_L$ gauge interactions, fermion content, and Higgs sector of the SM are mirrored, and the generalized parity symmetry constrains the new gauge, Yukawa, and Higgs quartic couplings  to match between the SM and mirror sectors. %The isosinglet fermion fields can be given parity-preserving vectorlike masses, while 
Parity is softly broken by mass splittings in the Higgs sector and by nonhermitian Dirac masses for the isosinglet fermions. 

As a consequence of the generalized parity $\bar\theta$ vanishes at tree level, but even more striking is the fact that the one loop corrections also vanish~\cite{Babu:1989rb}. This is a consequence of the limited new flavor structure in the model. In less minimal models, for example replacing the soft breaking of the Babu-Mohapatra model by a spontaneous parity-breaking sector, corrections may arise already at one loop, requiring certain couplings to be very small~\cite{Albaid:2015axa}.\footnote{We focus here on models where hypercharge is not mirrored. Radiative corrections to $\bar\theta$ can be much smaller in some models where the entire electroweak gauge group is doubled~\cite{barrsenjanovic}. These models are interesting, but also more phenomenologically constrained by the presence of a massless dark photon and stable, fractionally charged hadrons.}

Corrections to $\bar\theta$ in the Babu-Mohapatra model start at two loop order. These were estimated in~\cite{Babu:1989rb} to be of order $\bar\theta \lesssim 10^{-12}$ based on a qualitative analysis of a small subset of the Feynman graphs, and also based on an argument that the corrections should decouple like $v/v'$, where $v$ is the weak scale and $v'$ is the mirror weak scale. If the corrections are indeed this small, two conclusions can be drawn. First, the model successfully addresses the strong $CP$ problem, since searches for a neutron electric dipole moment (EDM) imply an upper limit of $\bar\theta\lesssim 10^{-10}$. Second, as recently computed in~\cite{Craig:2020bnv}, one-loop contributions the neutron EDM from dimension-6 operators could be comparable to the contribution from a $\bar\theta$-term of order $10^{-12}$. Thus the predicted spectrum of dipole moments in the model would depend on the detailed contributions both from the dimension-4 $\bar\theta$-term and a variety of dimension-6 operators.

However, the radiative corrections in this model are deserving of further scrutiny. Because $\bar\theta$ is a marginal coupling,  threshold corrections to it need not decouple as the new degrees of freedom are made heavy, as was is assumed in \cite{Babu:1989rb}.  We will see below that this is indeed the case, and there are potentially large corrections that reflect  general effective field theory (EFT) expectations. Consequently, the model does not automatically solve the strong $CP$ problem. 

It is possible, however, that the corrections are sufficiently small in some subsets of parameter space. Then on general EFT grounds one expects (as in any UV solution to the strong $CP$ problem) that  the pattern of low-energy EDM observables  is dominated by a small but nonzero $\bar\theta$ term rather than higher dimensional operators~\cite{deVries:2018mgf,deVries:2021sxz}.    Dimension-6 operators such as quark electric or chromo-electric dipole moments induce quadratically divergent corrections to $\bar \theta$, indicating the presence of parametrically unsuppressed (i.e. not suppressed by $v/\Lambda$) threshold corrections at the scale $\Lambda$ where the UV theory is matched to the SM-EFT. In contrast, the contribution from higher-dimension $CP$ violating operators is expected to have additional suppression by powers of $v/\Lambda$.  That is, within  UV solutions to strong $CP$, one expects hadronic EDMs to be dominated by the $\bar \theta$ term, rather than a model-dependent mixture of operators, as might be the case if $\bar\theta$ were indeed suppressed by $v/v'$~\cite{Babu:1989rb,Craig:2020bnv}.\footnote{One may also conclude from this argument that if a pattern of EDMs is observed that is inconsistent with a pure $\bar\theta$ term, there are likely large threshold corrections to $\bar\theta$ which can only be relaxed by an infrared solution to the strong $CP$ problem~\cite{deVries:2018mgf,deVries:2021sxz}.}

In this work we perform a more careful analysis of the two loop corrections to $\bar\theta$ in the softly-broken parity model. For simplicity, we assume a hierarchy among the weak scale $v$ and the scales associated with new fields, and we perform the calculation in the symmetric phase with positive scalar mass parameters. The leading correction is insensitive to the scale hierarchies, so it is accurately captured by the symmetric phase computation. These approximations are sufficient to demonstrate the basic physical points of the previous paragraph. Our main general results are given in (\ref{eq:ImTrYu}, \ref{eq:ImTrYd}) below, and our estimates for their magnitude at simple points in parameter space are in Section \ref{sec:magnitude}. We find that the corrections to $\bar\theta$ are in the range $10^{-3}\mbox{--}10^{-2}$ multiplying model-dependent mixing angles and phases.  Thus, viability of the model requires rather small angles and phases. On one hand the requirement of small parameters might be viewed as at odds with the original motivation of the model. On the other hand, the ``necessary smallness" is somewhat better than simply setting $\bar\theta=10^{-10}$ by hand, so more optimistically one might view this result as a datum pointing to the correct parameter space.
In the latter case we can conclude that the natural prediction for $\bar\theta$ in these models is not much smaller than the present limit. Thus we can view them as a target for ongoing and near-term hadronic EDM experiments, with a spectrum dominated by contributions from $\bar\theta$. We conclude by discussing how ratios of EDMs can be used to confirm this hypothesis.

\section{Babu-Mohapatra Model}
We begin by reviewing the model of softly-broken generalized parity introduced in \cite{Babu:1989rb}, following notation similar to \cite{Craig:2020bnv}.  The model is based on enlarging the electroweak gauge group to
\begin{equation}
SU(2)_\text{L} \times SU(2)_\text{R} \times U(1)_{\hat{Y}},
\end{equation}
with gauge potentials denoted $W_\mu$, $W'_\mu$, and $B_\mu$, and doubling the Weyl fermion and Higgs degrees of freedom, with charge assignments given by
\begin{gather}
\begin{aligned}
Q &\sim{} (\textbf{2}, 1, 1/6) & Q'^\dag &\sim{} (1, \textbf{2}, 1/6)\\
U^\dag &\sim{} (1, 1, 2/3) & U' &\sim{} (1, 1, 2/3)\\
D^\dag &\sim{} (1, 1, -1/3) & D' &\sim{} (1, 1, -1/3)\\
H &\sim{} (\textbf{2}, 1, 1/2) & H^{\prime*} &\sim{} (1, \textbf{2}, 1/2)\,.
\end{aligned}
\end{gather}
Generalized parity exchanges the unprimed and primed fields:
\begin{gather}
\begin{aligned}
W^\mu &\leftrightarrow{} W'_\mu\\
Q,U,D &\leftrightarrow{} Q'^\dag, U'^\dag, D'^\dag\\
H &\leftrightarrow{} H'^*\,,
\end{aligned}
\end{gather}
with all other fields transforming conventionally under parity inversion.
The tree-level Yukawa Lagrangian consistent with generalized parity invariance is
\begin{multline}\label{eq:defYukawa}
\mathcal{L} = - \begin{pmatrix}Q'H' & U\end{pmatrix}
\begin{pmatrix} 0 & Y_u \\ Y^\dag_u & \mathcal{M}_u \end{pmatrix}
\begin{pmatrix}Q H \\ U'\end{pmatrix}\\
- \begin{pmatrix}Q'{H'^*} & D\end{pmatrix}
\begin{pmatrix} 0 & Y_d \\ Y^\dag_d & \mathcal{M}_d \end{pmatrix}
\begin{pmatrix}Q H^* \\ D'\end{pmatrix} + \text{c.c.}\,,
\end{multline}
where $Y_{u,d}$ and $\mathcal{M}_{u,d}$ are $3\times3$ matrices in flavor space.  Collectively, we denote the $6\times6$ matrices appearing in (\ref{eq:defYukawa}) as $\mtxY_u^{(0)}$ and $\mtxY_d^{(0)}$.  The invariant $CP$-violating parameter $\bar\theta$ is the sum of the QCD vacuum angle $\theta_\text{QCD}$ and the phase of the determinant of the Yukawa matrix product
\begin{equation}\label{eq:invTheta}
\bar\theta = \theta_\text{QCD} + \argdet(\mtxY_u \mtxY_d)\,.
\end{equation}
Tree-level generalized parity requires $\theta_\text{QCD} = 0$, and zeroes in the upper-left block of $\mtxY_u^{(0)}$ and $\mtxY_d^{(0)}$ given in (\ref{eq:defYukawa}).  Consequently $\argdet(\mtxY_u^{(0)} \mtxY_d^{(0)}) = 0$, providing a tree-level solution to the strong $CP$ problem. 

The Higgs potential is given by
\begin{multline}
V(H,H') = \mu_L^2 |H|^2 + \mu_R^2 |H'|^2 + \lambda_p |H|^2 |H'|^2 \\
 + \lambda \big(|H|^4 + |H'|^4 \big)
\end{multline}
with $\mu_L^2,\,\mu_R^2 < 0$ so that both Higgs fields $H$ and $H'$ acquire vacuum expectation values given by $v$ and $v'$, respectively.  Although generalized parity requires $\mu_L^2 = \mu_R^2$, phenomenological viability requires parity to be broken in the vacuum.  In this model, generalized parity is allowed to be broken softly via
\begin{subequations}
\begin{align}
\label{eq:softPV1} &\mathcal{M}_{u,d} \neq \mathcal{M}_{u,d}^\dag \,,\\
\label{eq:softPV2} \text{and }&\mu_L^2 \neq \mu_R^2\,.
\end{align}
\end{subequations}
These sources of soft parity breaking radiatively generate a finite renormalization of the effective vacuum parameter $\bar\theta$ through the Yukawa matrices $\mtxY_u$ and $\mtxY_d$.  One-loop shifts to $\bar\theta$ were shown to vanish in \cite{Babu:1989rb}, although two-loop shifts are expected to be nonzero.

Numerous phenomenological aspects of the model were recently studied in Ref.~\cite{Craig:2020bnv}. 
As explained in their work, the hierarchy
\begin{equation}\label{eq:seesawlimit}
\mu_L   \ll  \mu_R  \ll   |\mathcal{M}_{u,d}|
\end{equation}
achieves SM quark masses near their observed value by the seesaw mechanism (apart from the top quark mass) while maintaining a moderate level of fine-tuning in the scalar sector.  Obtaining the correct top quark mass requires special treatment since perturbativity of the top Yukawa couplings forbids a large hierarchy between $v$, $v'$ and the relevant elements of $\mathcal{M}_u$.  Perturbativity of the bottom Yukawa requires the seesaw scale not to be vastly larger the parity breaking scale, $v'/|\mathcal{M}|\gtrsim 10^{-2}$.

\section{$\bar\theta$ at two loops}
We are interested in computing two-loop contributions to $\bar\theta$ to lowest non-vanishing order in the seesaw limit \eqref{eq:seesawlimit}.  To simplify this task as much as possible, we carry out our computation in the symmetric phase, taking $\mu_L^2$ and $\mu_R^2$ positive.  This approach gives the correct leading seesaw behavior in terms of the underlying model parameters, provided the result is IR finite as $v, v'\rightarrow0$. We discuss which subleading corrections receive additional contributions in the broken phase further below.

Writing the tree-, one-, and two-loop contributions to the effective $6\times6$ matrices as $\mtxY_q = \mtxY_q^{(0)} + \mtxY_q^{(1)} + \mtxY_q^{(2)}$ in each sector $q=\{u, d\}$, the expansion of $\argdet(\mtxY_q)$ through two-loop order entering into \eqref{eq:invTheta} reads 
\begin{multline}\label{eq:argdetExp}
\argdet(\mtxY_q) =  \argdet(\mtxY_q^{(0)}) + \imtr (\frac{1}{\mtxY_q^{(0)}}\mtxY_q^{(1)})\\
 + \imtr (\frac{1}{\mtxY_q^{(0)}}\mtxY_q^{(2)}) - \frac{1}{2} \imtr (\frac{1}{\mtxY_q^{(0)}}\mtxY_q^{(1)}\frac{1}{\mtxY_q^{(0)}} \mtxY_q^{(1)})\,.
\end{multline}
The first two terms are the tree-level and one-loop contributions that vanish by generalized parity and as established in \cite{Babu:1989rb}, respectively.  The two terms in the second line are the 1PI irreducible and reducible contributions respectively at the two-loop order.

For convenience, we transform to a basis in which the initially nonhermitian isosinglet mass matrices $\mathcal{M}_{u,d}$ are diagonal and real.  On account of \eqref{eq:softPV1} this requires separate unitary transformations for each of the four isosinglet quarks $\{U^{(\prime)}, D^{(\prime)}\}\rightarrow\{\tilde{U}^{(\prime)}, \tilde{D}^{(\prime)}\}$.  In this basis, the Yukawa Lagrangian (\ref{eq:defYukawa}) becomes
\begin{multline}\label{eq:defYukawaIsodiag}
\mathcal{L} = - \begin{pmatrix}Q'H' & \tilde{U}\end{pmatrix}
\begin{pmatrix} 0 & \tilde{Y}'_u \\ \tilde{Y}^\dag_u & \mathcal{M}_u^\text{diag} \end{pmatrix}
\begin{pmatrix}Q H \\ \tilde{U}'\end{pmatrix}\\
- \begin{pmatrix}Q'{H'^*} & \tilde{D}\end{pmatrix}
\begin{pmatrix} 0 & \tilde{Y}'_d \\ \tilde{Y}^\dag_d & \mathcal{M}_d^\text{diag} \end{pmatrix}
\begin{pmatrix}Q H^* \\ \tilde{D}'\end{pmatrix} + \text{c.c.}\,,
\end{multline}
and the soft parity breaking source \eqref{eq:softPV1} is now encoded in $\tilde{Y}_{u,d} \neq \tilde{Y}'_{u,d}$. Note that this is not a hard breaking of parity, since it can always be moved back to the dimensionful isosinglet mass matrices by a change of basis.

\begin{figure}[t]
\centering
\includegraphics[width=3.4cm]{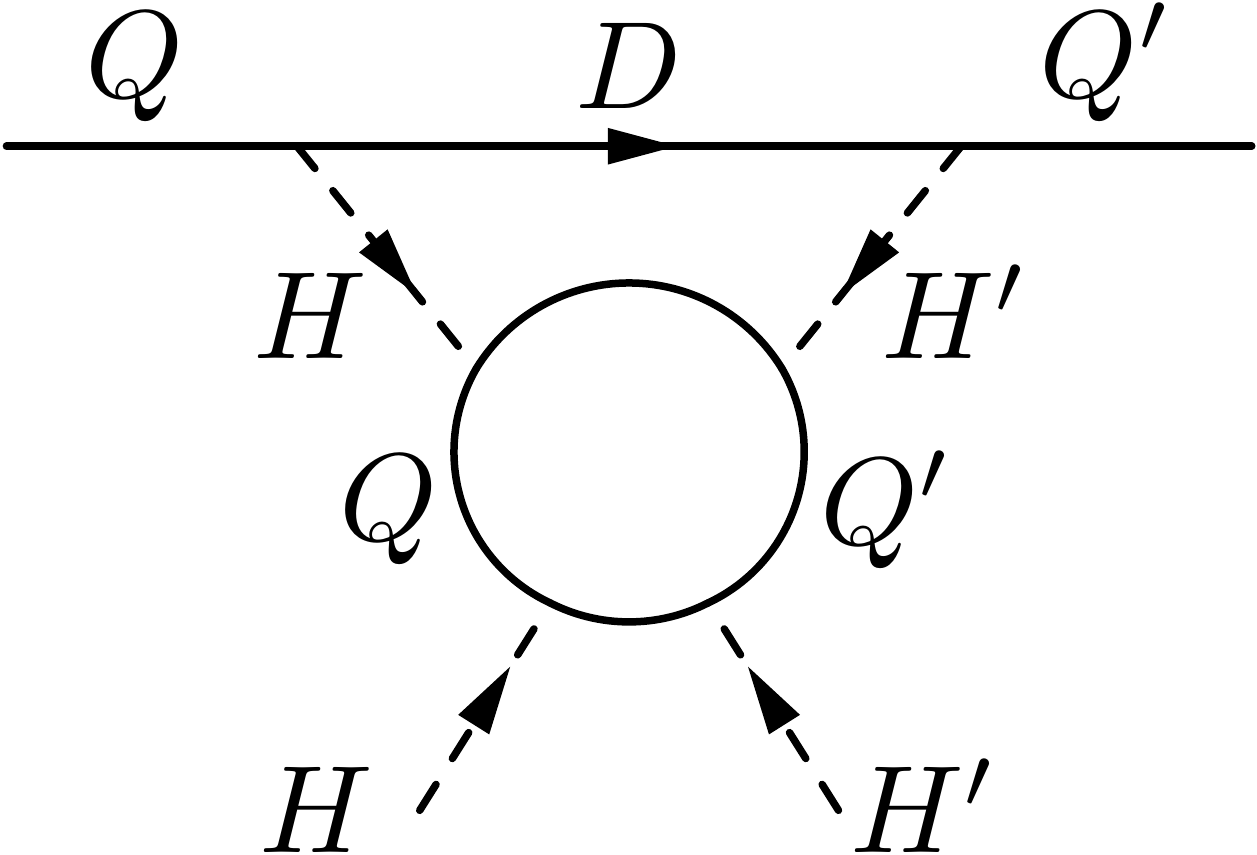}\enspace\enspace
\includegraphics[width=3.4cm]{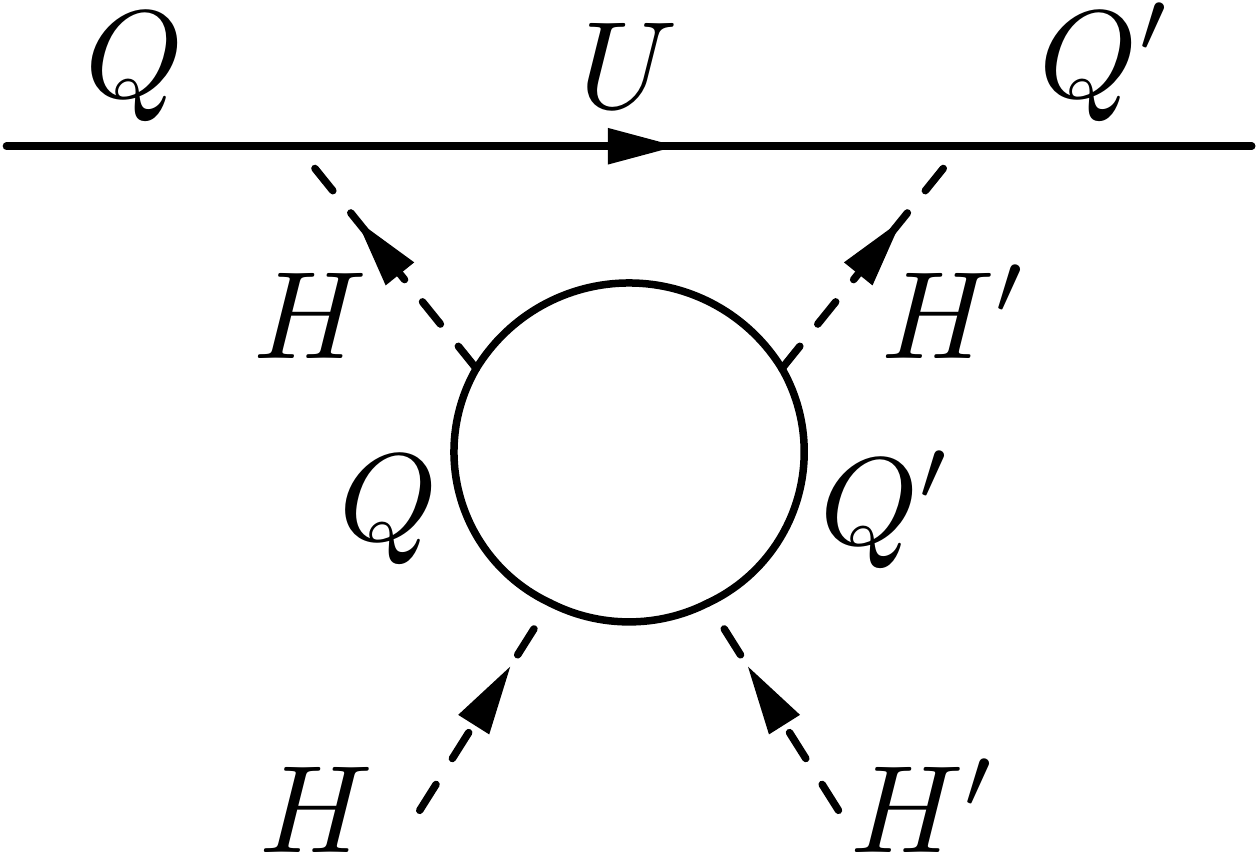}\\[4mm]
\includegraphics[width=3.4cm]{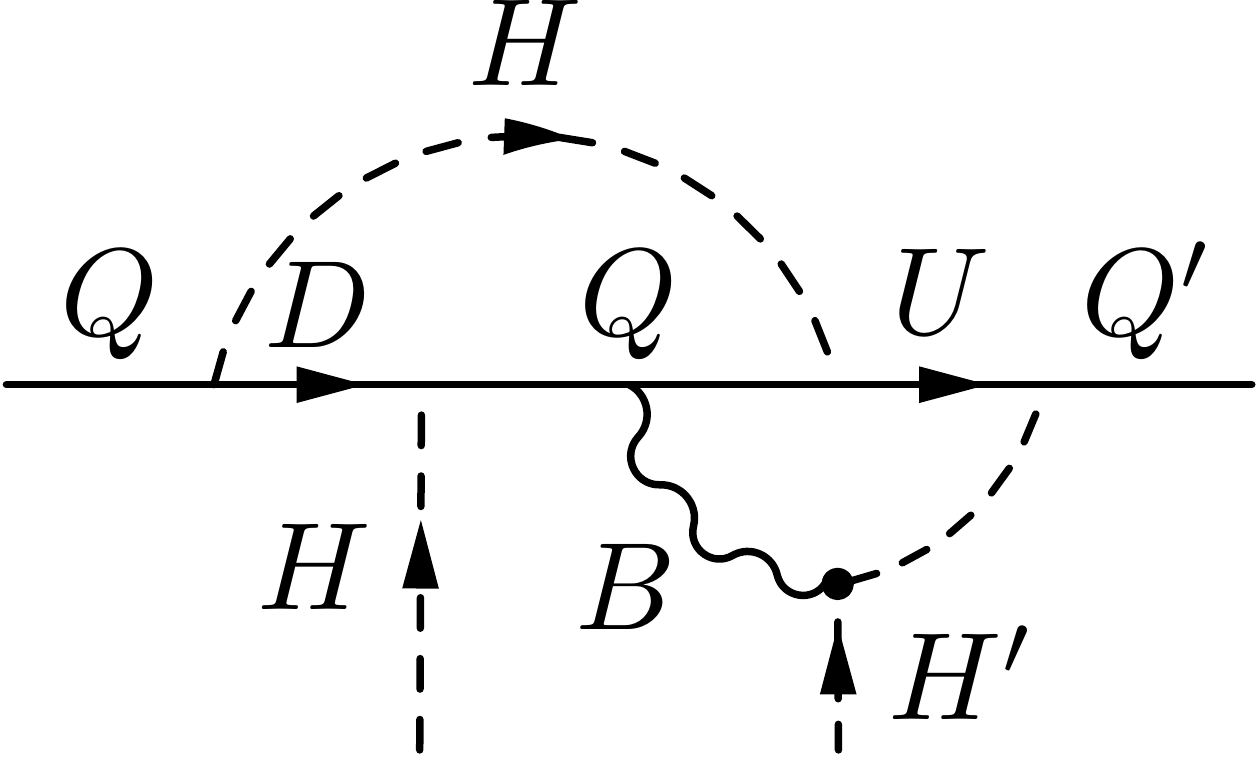}\enspace\enspace
\includegraphics[width=3.4cm]{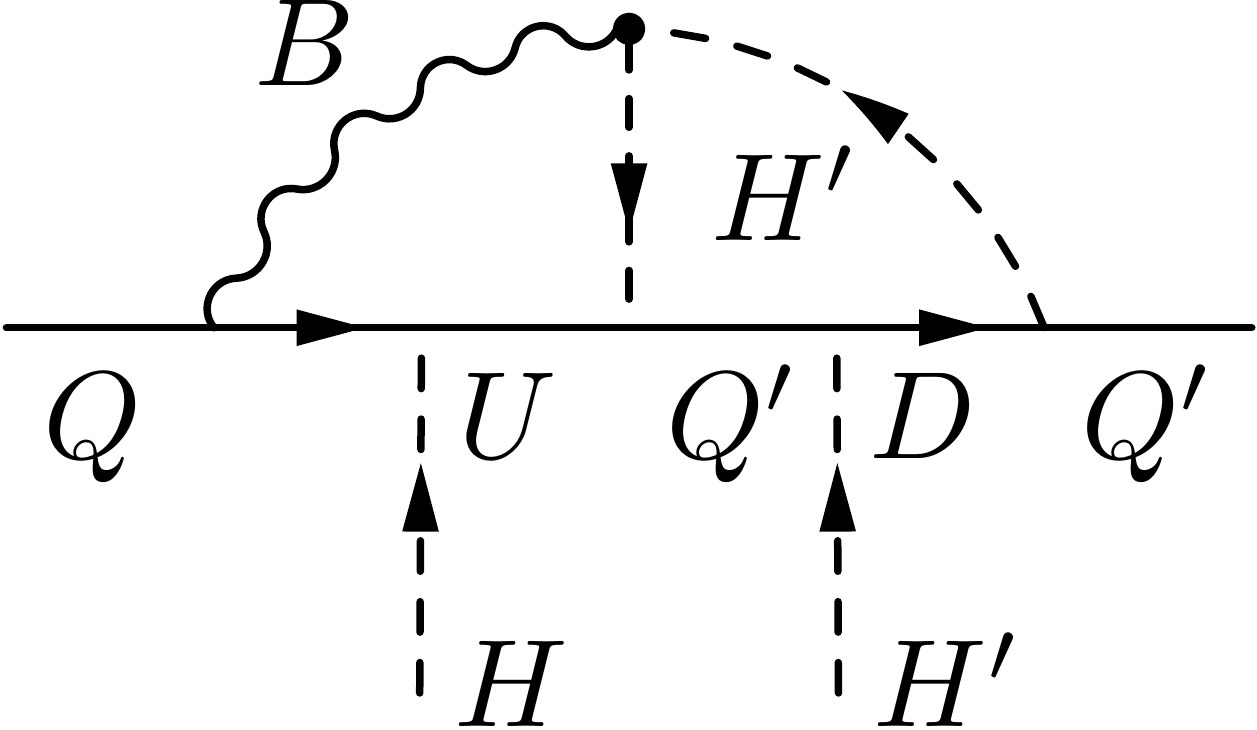}\\[4mm]
\includegraphics[width=3.4cm]{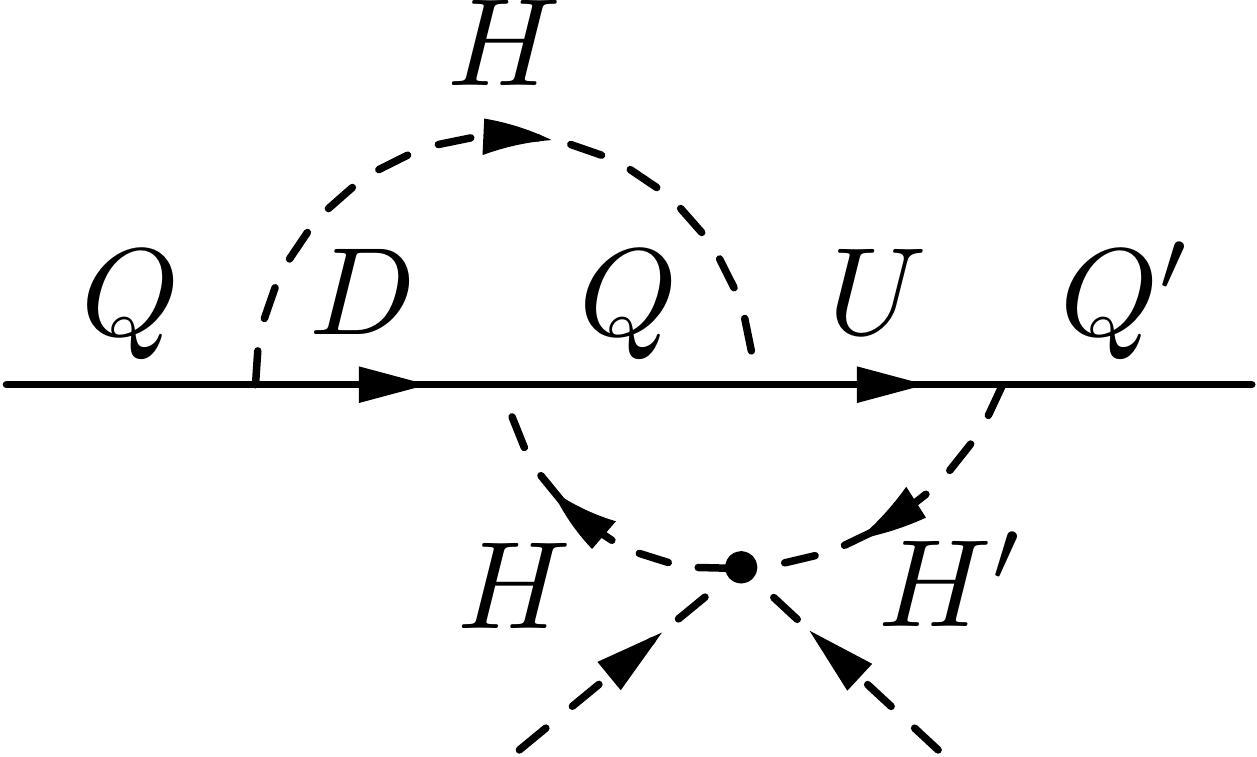}\enspace\enspace
\includegraphics[width=3.4cm]{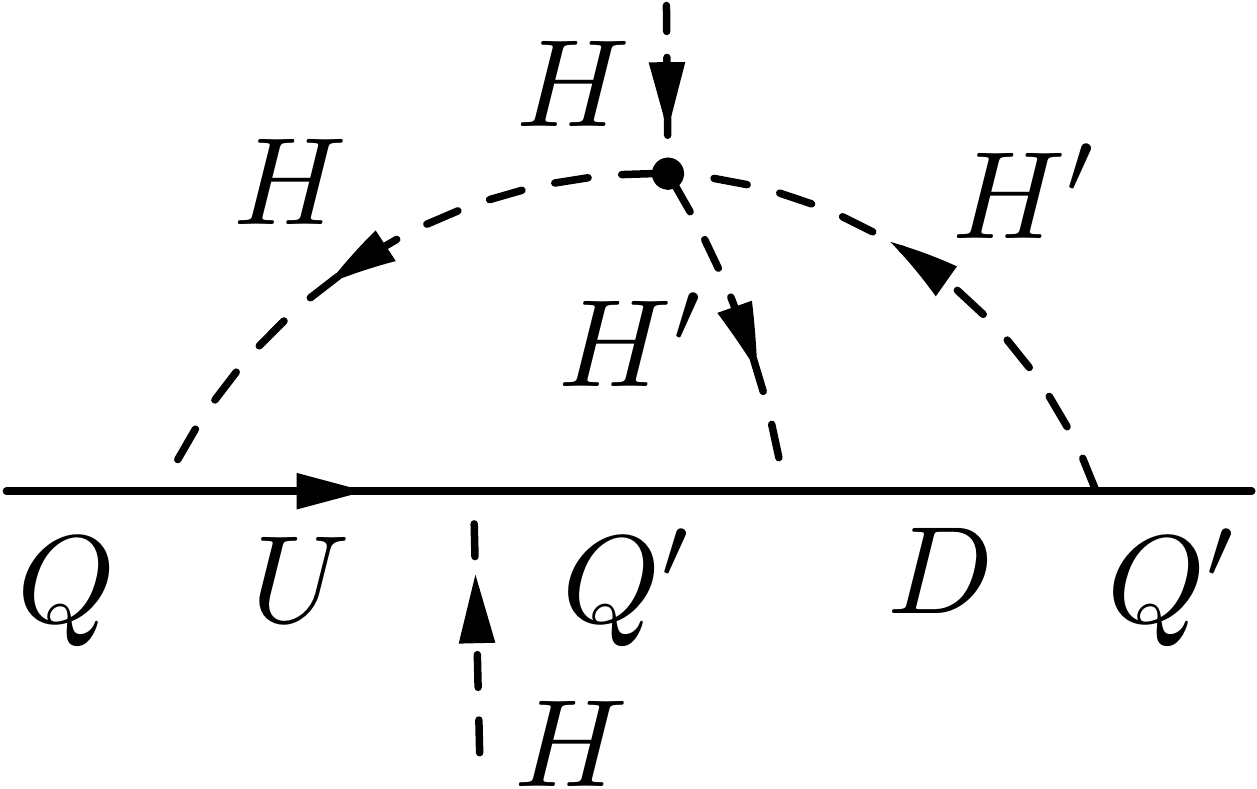}\\[4mm]
\includegraphics[width=3.4cm]{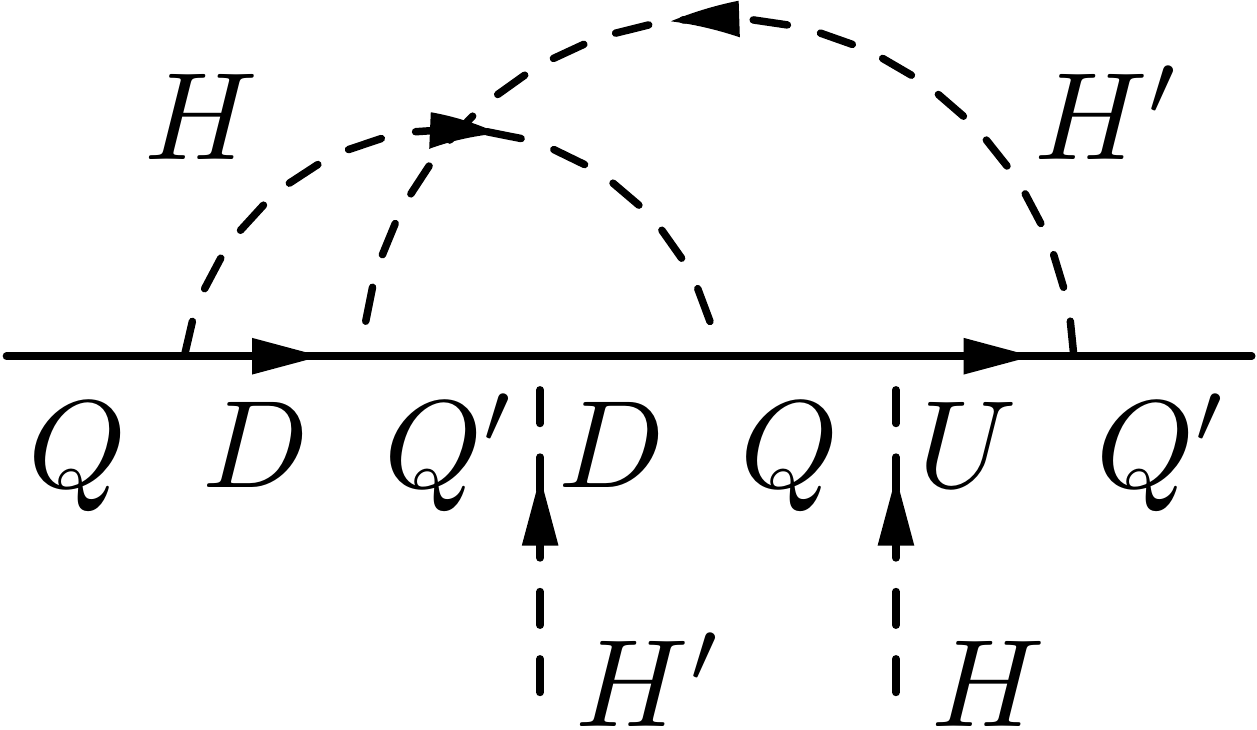}\enspace\enspace
\includegraphics[width=3.4cm]{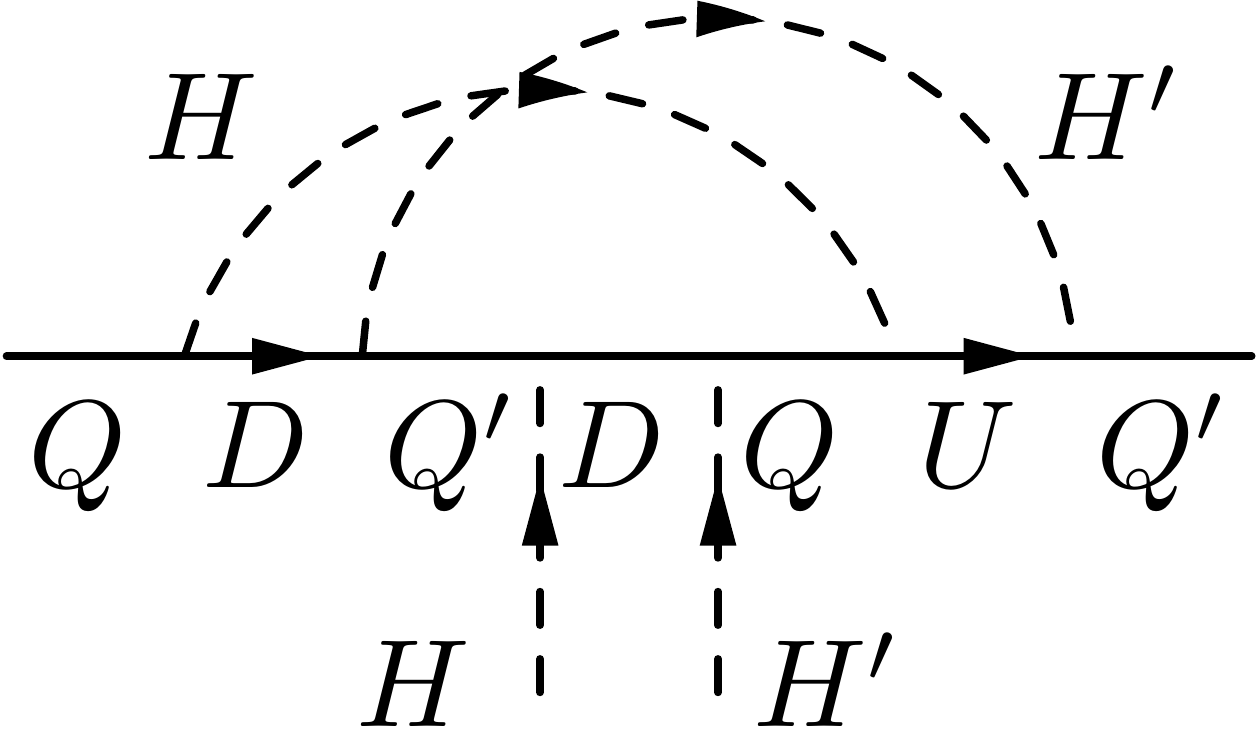}\\[4mm]
\includegraphics[width=3.4cm]{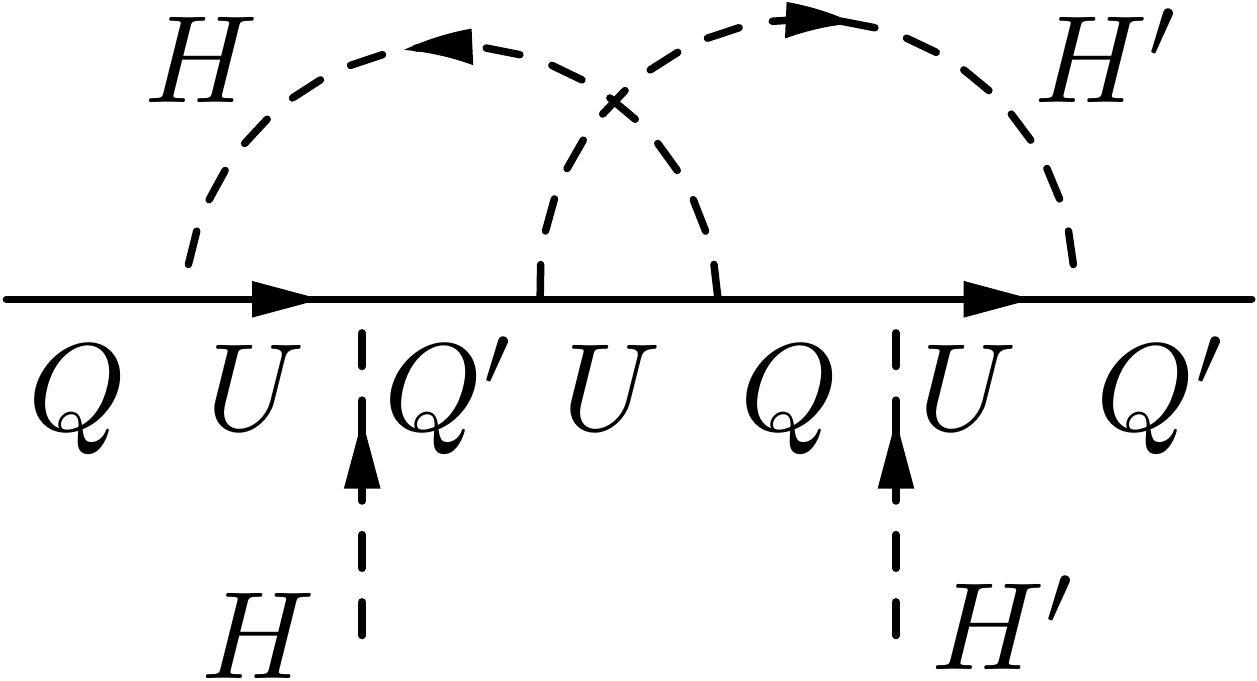}
\caption{
Representative 4-point 2-loop diagrams that contribute to loop functions $A$ and $B$ (first row, left and right, respectively), $F_u$ (second row), $G$ (third row), $H$ (fourth row) and $J$ (fifth row) in \eqref{eq:ImTrYu}.}
\label{fig:FeynmanDiagrams}
\end{figure}

We implemented Feynman rules of this model in the symmetric phase and in the isosinglet mass basis into \textsc{FeynArts} \cite{Hahn:2000kx}.   We generated all 2-, 3-, and 4-point Feynman diagrams that contribute to $\mathbb{Y}_q^{(1)}$ and $\mathbb{Y}_q^{(2)}$ in the Fermi $\xi$-gauges, and evaluated them with an in-house \textsc{Mathematica} program.  Diagrams without an internal Higgs $H$ or $H'$ line do not give an imaginary part.  Similarly, diagrams involving internal $W_\mu$ or $W'_\mu$ gauge fields also vanish.  Although diagrams with an internal gluon generate non-vanishing imaginary parts, their contributions cancel out upon adding together the irreducible and reducible terms.  In terms of the diagonal elements of the isosinglet mass matrix
\begin{gather}\begin{aligned}
  \mathcal{M}^\text{diag}_u &=  \text{diag}(M_{u1}, M_{u2}, M_{u3})\\
  \mathcal{M}^\text{diag}_d &=  \text{diag}(M_{d1}, M_{d2}, M_{d3}),
\end{aligned}\end{gather}
we find the following result:

\begin{widetext}
\begin{align}\label{eq:ImTrYu}
\nonumber \argdet(\mtxY_u) = {} &\frac{1}{(16\pi^2)^2}\times \\
\nonumber \text{Im}\Bigg\{
\sum_{i,j,k,\ell}^3 & (\tilde{Y}^{\prime\dag}_d \tilde{Y}'_u)_{ij} (\tilde{Y}_u^\dag \tilde{Y}_d)_{ji}  (\tilde{Y}_u^{-1} \tilde{Y}_d)_{k\ell} (\tilde{Y}_d^{\prime\dag} \tilde{Y}_u^{\prime\dag-1})_{\ell k} A(\mu_L^2, \mu_R^2, M_{di}, M_{uj}, M_{uk},M_{d\ell})\\
\nonumber + \sum_{i,j,k}^3 & \Big[(\tilde{Y}_u^{\prime\dag} \tilde{Y}_u^\prime)_{jk} (\tilde{Y}_u^\dag \tilde{Y}_u)_{kj}  B(\mu_L^2, \mu_R^2, M_{ui},M_{uj},M_{uk})+ (\tilde{Y}_d^\dag \tilde{Y}_d)_{jk} (\tilde{Y}_d^{\prime\dag} \tilde{Y}_d^\prime)_{kj} B(\mu_L^2, \mu_R^2, M_{ui},M_{dj},M_{dk})\Big]\\
\nonumber + g'^2\sum_{i,j = 1}^3 &\Big[ (\tilde{Y}_u^\dag \tilde{Y}_d)_{ij} (\tilde{Y}_d^\dag \tilde{Y}_u^{\dag-1})_{ji}   F_u(\mu_L^2, \mu_R^2, M_{ui},M_{dj}) - (\tilde{Y}_u'^\dag \tilde{Y}_d')_{ij}(\tilde{Y}^{\prime\dag}_d \tilde{Y}_u^{\prime\dag-1})_{ji}   F_u(\mu_R^2, \mu_L^2, M_{ui},M_{dj})\Big]\\
\nonumber +\lambda_p \sum_{i,j = 1}^3 &\Big[(\tilde{Y}_u^\dag \tilde{Y}_d)_{ij} (\tilde{Y}_d^\dag \tilde{Y}_u^{\dag-1})_{ji} G(\mu_L^2, \mu_R^2, M_{ui},M_{dj}) - (\tilde{Y}_u'^\dag \tilde{Y}_d')_{ij} (\tilde{Y}^{\prime\dag}_d \tilde{Y}_u^{\prime\dag-1})_{ji}    G(\mu_R^2, \mu_L^2, M_{ui},M_{dj})\Big]\\
\nonumber + \sum_{i,j,k = 1}^3 &\Big[(\tilde{Y}_u^\dag \tilde{Y}_d)_{ij} (\tilde{Y}_d^{\prime\dag} \tilde{Y}'_d)_{jk} (\tilde{Y}^\dag_d \tilde{Y}_u^{\dag-1})_{ki}   H(\mu_L^2, \mu_R^2, M_{ui},M_{dj},M_{dk})\\
\nonumber &\hspace{3.5cm}- (\tilde{Y}_u'^\dag \tilde{Y}_d')_{ij}(\tilde{Y}_d^\dag \tilde{Y}_d)_{jk} (\tilde{Y}_d^{\prime\dag} \tilde{Y}_u^{\prime\dag-1})_{ki}   H(\mu_R^2,\mu_L^2,M_{ui},M_{dj},M_{dk})\Big]\\
 + \sum_{i,j}  & (\tilde{Y}_u^{\prime\dag} \tilde{Y}'_u)_{ij} (\tilde{Y}_u^\dag \tilde{Y}_u)_{ji} J(\mu_L^2, \mu_R^2, M_{ui},M_{uj})\Bigg\}\,,\\
\label{eq:ImTrYd} \argdet(\mtxY_d) = {}&\argdet(\mtxY_u)\Big|_{u\leftrightarrow d}\,.
\end{align}
\end{widetext}
The loop functions denoted $A$, $B$, $F_{u,d}$, $G$, $H$, and $J$, are real and dimensionless functions of Higgs masses $\mu_{L,R}^2$ and isosinglet quark masses $(M_{u,d})_i$.  Representative Feynman diagrams are shown in Fig.~\ref{fig:FeynmanDiagrams}.   The corresponding result in the down sector \eqref{eq:ImTrYd} is obtained from the up sector result \eqref{eq:ImTrYu} by exchanging Yukawa matrices $\tilde{Y}_u^{(\prime)}\leftrightarrow \tilde{Y}_d^{(\prime)}$, exchanging isosinglet quark mass arguments $M_{ui}\leftrightarrow M_{di}$, and by replacing the loop function $F_u \rightarrow F_d$, which depends on quark hypercharges.  We perform an expansion of the loop functions according to \eqref{eq:seesawlimit} and consistent with our calculation in the symmetric phase.  Notably, we find that the expansion starts at zeroth order in $\mu_{L,R}^2/|\mathcal{M}_{u,d}|^2$, confirming our expectation that parity violation does not decouple from $\bar\theta$  as $|\mathcal{M}_{u,d}|\rightarrow\infty$, in contrast to the claim of \cite{Babu:1989rb}.  This may be regarded as a threshold correction to $\bar\theta$ at the scale $M$ from the non-hermiticity of the isosinglet mass matrices. As validation of our results, we have checked that our final results are independent of gauge-fixing parameters, and are UV and IR finite to lowest nonvanishing order in (\ref{eq:seesawlimit}).

Finally, we readily verify that \eqref{eq:ImTrYu} vanishes in the limit that the sources of generalized parity violation in (\ref{eq:softPV1},\ref{eq:softPV2}) vanish.  
First, putting $\mathcal{M}_{u,d} = \mathcal{M}_{u,d}^\dag$ implies $\tilde{Y}_{u,d} = \tilde{Y}'_{u,d}$ in the isosinglet mass basis.  In this case, the product of Yukawa matrices proportional to $A$, $B$ and $J$ in \eqref{eq:ImTrYu} become purely real, so these terms drop out.  The flavor matrices proportional to $F_{u,d}$, $G$, and $H$ combine to yield
\begin{align}\label{eq:HermImTrYu}
& \hspace{-1.67cm}\nonumber \argdet(\mtxY_u)\big|_{\mathcal{M}=\mathcal{M}^\dag} = \frac{1}{(16\pi^2)^2}\text{Im}\Big\{ \\
\nonumber 
g'^2\sum_{i,j = 1}^3 &\Big[ (\tilde{Y}_u^\dag \tilde{Y}_d)_{ij} (\tilde{Y}_d^\dag \tilde{Y}_u^{\dag-1})_{ji}   \bar{F}_u(\mu_L^2, \mu_R^2, M_{ui},M_{dj}) \Big]\\
\nonumber +\lambda_p \sum_{i,j = 1}^3 &\Big[(\tilde{Y}_u^\dag \tilde{Y}_d)_{ij} (\tilde{Y}_d^\dag \tilde{Y}_u^{\dag-1})_{ji} \bar{G}(\mu_L^2, \mu_R^2, M_{ui},M_{dj}) \Big]\\
\nonumber + \sum_{i,j,k = 1}^3 &\Big[(\tilde{Y}_u^\dag \tilde{Y}_d)_{ij} (\tilde{Y}_d^\dag \tilde{Y}_d)_{jk} (\tilde{Y}^\dag_d \tilde{Y}_u^{\dag-1})_{ki}\times\\
&\hspace{1.5cm}\bar{H}(\mu_L^2, \mu_R^2, M_{ui},M_{dj},M_{dk}) \Big\}\,,\\
\label{eq:HermImTrYd} & \hspace{-1.67cm} \argdet(\mtxY_d)\big|_{\mathcal{M}=\mathcal{M}^\dag} = \argdet(\mtxY_u)|_{\mathcal{M}=\mathcal{M}^\dag}\Big|_{u\leftrightarrow d}\,,
\end{align}
where $\bar{F}_{u,d}$, $\bar{G}$ and $\bar{H}$ are the antisymmetric combinations under $\mu_L^2 \leftrightarrow \mu_R^2$
\begin{multline}
\bar{F}_u(\mu_L^2, \mu_R^2, M_{ui},M_{dj}) = \\
 F_u(\mu_L^2, \mu_R^2, M_{ui},M_{dj}) - F_u(\mu_R^2, \mu_L^2, M_{ui},M_{dj})\,,
\end{multline}
\emph{etc.}, and therefore are proportional to the remaining source of parity breaking $(\mu_R^2-\mu_L^2)/M^2$.  Taking $\mu_L^2 \rightarrow \mu_R^2$, these remaining loop functions vanish, so that $\argdet(\mtxY_u) = \argdet(\mtxY_d) = 0$ as required in the absence of any source of generalized parity breaking.

\section{Magnitude of the corrections}
\label{sec:magnitude}
Now we would like to make numerical estimates for the size of the two loop shift in $\bar\theta$. To do so is complicated by the need to fit a subset of the microscopic parameters in $\tilde{Y}^{(\prime)}_{u,d}$ and $\mathcal{M}_{u,d}^\text{diag}$ in (\ref{eq:defYukawaIsodiag}) to the known SM quark masses and CKM elements. To simplify matters and to obtain order-of-magnitude estimates, we make certain assumptions for the spectrum following the approach of Ref.~\cite{Craig:2020bnv}. First, we take the isosinglet mass matrices to be degenerate, apart from the isosinglet top mass, which we set to zero,
\begin{gather}\label{eq:Massumpt}
\begin{aligned}
 \mathcal{M}_u^\text{diag} &=  \text{diag}(M, M, 0)\\
 \mathcal{M}_d^\text{diag} &=  \text{diag}(M, M, M)\, .
\end{aligned}
\end{gather}
Second, we assume that in this basis, various top quark mixing terms vanish
\begin{gather}
\begin{gathered}
(\tilde{Y}_u)_{31} =(\tilde{Y}_u)_{32}=(\tilde{Y}_u)_{13}=(\tilde{Y}_u)_{23} = 0\\
(\tilde{Y}'_u)_{31} =(\tilde{Y}'_u)_{32}=(\tilde{Y}'_u)_{13}=(\tilde{Y}'_u)_{23} = 0\,.
\end{gathered}
\end{gather}
Finally, we diagonalize $\tilde{Y}^{(\prime)}_d$ and the remaining elements of $\tilde{Y}^{(\prime)}_u$.  For consistency with the chosen spectrum in \eqref{eq:Massumpt}, the primed matrices $\tilde{Y}'_d$ and $\tilde{Y}'_u$ may only differ by an overall phase relative to their unprimed counterparts. (In other words, the initial non-hermiticity of the $\mathcal{M}_{u,d}$ could be encoded in two overall phases.)   We write the rotation matrices as
\begin{gather}
\begin{aligned}
\tilde{Y}_u &= V_u^\dag\, \text{diag}(y_u, y_c, y_t)\,  U \\
\tilde{Y}_d &= V_d^\dag\, \text{diag}(y_d, y_s, y_b)\,  W \\
\tilde{Y}_u' &= V_u^\dag\, \text{diag}(y_u, y_c, y_t)\,  U e^{-i\varphi_{u}}\\
\tilde{Y}_d' &= V_d^\dag\, \text{diag}(y_d, y_s, y_b)\,  W e^{-i\varphi_{d}}\,,
\end{aligned}
\end{gather}
with the product $V_u V_d^\dagger \equiv V_\text{CKM}$ yielding the Cabbibo-Kobayashi-Maskawa matrix.

At this stage we have diagonalized each generation separately, but within each generation we must still diagonalize the mixing between SM quarks and their mirror states, and fit the former to the known values.  The spectrum is independent of the phases $\varphi_{u}$ and $\varphi_{d}$.  The results are simple in the seesaw approximation \eqref{eq:seesawlimit} and in the vanishing-isosinglet-top-mass assumption \eqref{eq:Massumpt},
\begin{equation}
\begin{aligned}\label{Yukawa}
y_q &\sim{} \sqrt{\frac{2m_q M}{v v'}}\,,\enspace q=\{u,d,c,s,b\}\\
y_t &\sim{} \frac{\sqrt{2}m_t}{v}\,,
\end{aligned}
\end{equation}
which we use to fit to the known quark masses.
For the scalar sector parameters, we assume $\lambda_p$ is negligibly small so that $\mu_L^2 \approx m_H^2/2$ and $\mu_R^2 \approx m_H^2 v'^2/2 v^2$, where $v=246\text{ GeV}$ and $m_H = 125\text{ GeV}$.

%To proceed further we require forms of the loop functions relevant for the spectrum of isosinglet masses chosen in \eqref{eq:Massumpt}.  In this case, the final arguments of the loop functions in \eqref{eq:ImTrYu} are either $M$ or $0$.  The asymptotic behavior of the loop functions in the seesaw limit are then as follows:
%\begin{gather}
%A(\mu_L^2, \mu_R^2,M,M,M,M) = -3\,\big(\pi^2-3\,\Phi(1)\big)\\
%B(\mu_L^2, \mu_R^2,M,M,M) = 6\ln\Big(\frac{M^2}{\mu_R^2}\Big)-9(\Phi(1)-2)
%\end{gather}
%%
%\begin{align}
%F_u(\mu_{L,R}^2, \mu_{R,L}^2, M, M) &= \frac{1}{6}\Big(\frac{\pi^2}{6}+\frac{1}{4}\Big)\\
%F_u(\mu_{L,R}^2, \mu_{R,L}^2, 0,M) &=  \frac{\pi^2}{6}-\frac{5}{6}\\
%F_d(\mu_{L,R}^2, \mu_{R,L}^2, M, M) &= \frac{5 \pi^2}{36}-\frac{13}{24}\\
%F_d(\mu_{L,R}^2, \mu_{R,L}^2, M,0) &= - \frac{7}{24}
%\end{align}
%%
%\begin{align}
%H(\mu_{L,R}^2, \mu_{R,L}^2, M,M,M) &= \frac{1}{4}(\pi^2-3\,\Phi(1))\\
%J(\mu_L^2, \mu_R^2,M,M) &= -\frac{1}{4}(2-\pi^2 + 3\, \Phi(1))
%\end{align}
%where $\Phi(1) \approx 2.344$.  All other mass combinations are suppressed by $\mu_R^2/M^2$. 

In this simplified parametrization, only the term proportional to loop function $A$ exhibits dependence on $\varphi_{u}$ and $\varphi_{d}$.  All others vanish, or are suppressed by a power of $(\mu_R^2-\mu_L^2)/M^2$.  The leading order seesaw behavior of loop function $A$ is
\begin{equation}
A(\mu_L^2, \mu_R^2,M,M,M,M) = -3\,\big(\pi^2-3\,\Phi(1)\big)\,,
\end{equation}
where $\Phi(1) \approx 2.344$, and is down by $\mu_R^2/M^2$ if there is a zero in any of the final four arguments.  There are many terms that depend on different combinations of quark Yukawa couplings and CKM elements.  We find the largest terms in the up and down sectors are
\begin{align}
\nonumber \argdet&(\mtxY_u) = \frac{3}{64\pi^4} \sin\big(2( \varphi_u{-}\varphi_d)\big) \big(\pi^2 - 3\,\Phi(1)\big)\times\\
\nonumber\Big\{&-\frac{ m_b m_t^2}{v^3}\frac{M}{v'}\Big(\frac{m_s}{m_u}|V_{us}|^2
+\frac{m_d}{m_u}|V_{ud}|^2\Big)|V_{tb}|^2\times\\
\nonumber &\hspace{2cm}\big(1-|U_{u3}|^2\big)\big(1-|U_{t3}|^2\Big)\\
&-\frac{m_b^2}{v^2}\frac{M^2}{v'^2} |V_{tb}|^4 (1-|U_{t3}|^2)^2 + \ldots\Big\}\,,
\end{align}
and
\begin{multline}
\argdet(\mtxY_d) = \frac{3}{64\pi^4} \sin\big(2( \varphi_u{-}\varphi_d)\big) \big(\pi^2 - 3\,\Phi(1)\big)\times\\
\Big\{\frac{ m_b m_t^2}{v^3}\frac{M}{v'}\Big(\frac{m_c}{m_s}|V_{cs}|^2
+\frac{m_c}{m_d}|V_{cd}|^2\Big)|V_{tb}|^2\times\\
\big(1-|U_{c3}|^2\big)\big(1-|U_{t3}|^2\Big)+\ldots\Big\}\,.
\end{multline}
The appearance of positive powers of $M/v'$ is due to the seesaw scaling of the Yukawas in \eqref{Yukawa}. Normalizing to $v'/M = 0.02$, which is near the perturbativity limit for $y_b$, yields
\begin{align}
\nonumber \argdet&(\mtxY_u) \approx{} \sin\big(2( \varphi_u{-}\varphi_d)\big)\times\\
\nonumber  \Big\{&-(\num{2.4e-3})\Big(\frac{0.02}{v'/M}\Big) \big(1-|U_{u3}|^2\big)\big(1-|U_{t3}|^2\Big) \\
&-(\num{1.0e-3})\Big(\frac{0.02}{v'/M}\Big)^2 (1-|U_{t3}|^2)^2\Big\}\,,
\end{align}
and
\begin{multline}
\argdet(\mtxY_d) \approx{} \sin\big(2( \varphi_u{-}\varphi_d)\big)\times\\
 (\num{1.5e-2})\Big(\frac{0.02}{v'/M}\Big) \big(1-|U_{c3}|^2\big)\big(1-|U_{t3}|^2\Big)\,.
\end{multline}
Even as we move away from the seesaw limit, this result suggests that the corrections are not smaller than $10^{-3}$ times mixing angles and phases.

Since the model does not solve strong $CP$ for generic mixings and a restricted set of soft parity breakings, it also will not solve strong $CP$ for a generic set of soft breaking parameters. However, one might hope that it could still function in a more specialized part of parameter space. To keep $\bar\theta$ small, it is clearly necessary to have other small numbers in the model. For the benchmark above it would be sufficient to have $|U_{t3}|\sim 0.9999$, with all other numbers of order one. Whether it is possible to find a satisfactory parameter space that is consistent with constraints, has nonhermitian isosinglet masses, and does not have additional fine tunings is an open question. To address it would require a much more detailed analysis than we perform here, including a complete computation in the broken phase.

Another possibility for the requisite small numbers is to restrict the soft parity breaking to the Higgs sector $\mu_L^2 \neq \mu_R^2$, requiring the isosinglet mass matrices $\mathcal{M}_u$ and $\mathcal{M}_d$ to be nearly hermitian.  Although this limit cannot be regarded as technically natural, the Higgs mass parameters do not contribute to the beta functions of the isosinglet fermion masses, and therefore this is a radiatively stable parameter manifold.  Our two-loop result for $\bar\theta$ in this limit reduces to the formulae in \eqref{eq:HermImTrYu} and \eqref{eq:HermImTrYd}, plus some contributions involving additional insertions of $H^{\prime\dag}H'$ that we have not computed here.  Due to these missing contributions, the evaluation below is somewhat schematic, but it can still serve as an order-of-magnitude estimate.  In our parametrization \eqref{eq:Massumpt} with degenerate isosinglet down quarks, $\argdet(\mtxY_d)\big|_{\mathcal{M}=\mathcal{M}^\dag}$ is zero in the down quark sector since all flavor matrix products are hermitian and do not give an imaginary part.

By unitarity of the rotation matrices $U$ and $W$, loop functions enter in the various terms of $\argdet(\mtxY_u)\big|_{\mathcal{M}=\mathcal{M}^\dag}$ as differences.  The asymptotic behaviors of these differences are
%\begin{align}
%\bar{F}_u(M,M) &={} \frac{\mu_R^2}{4M^2}\Big(\frac{7\pi^2}{18}-1\Big)\\
%\bar{F}_u(0,M) &={} \frac{3\mu_R^2}{4M^2}\Big(\ln\big(\frac{M^2}{\mu_R^2}\big) - \frac{\pi^2}{18}\Big)\\
%\bar{F}_d(M,M) &={} \frac{-3\mu_R^2}{4M^2}{\Big(\frac{\pi^2}{54}-1\Big)}\\
%\bar{F}_d(M,0) &={} \frac{3\mu_R^2}{4M^2}{\Big(\ln\big(\frac{M^2}{\mu_R^2}\big) - \frac{5\pi^2}{18}+2\Big)}
%\end{align}
%\begin{align}
%\bar{G}(M,M) &={} -\frac{\mu_R^2}{2M^2}\\
%\bar{G}(0,M) &={} \frac{-\mu_R^2}{4 M^2}\Big(\ln^2\big(\frac{M^2}{\mu_R^2}\big) - \ln\big(\frac{M^2}{\mu_R^2}\big)+ \frac{2 \pi^2}{3} - \frac{5}{2}\Big)\\
%\bar{G}(M,0) &={} -\frac{\mu_R^2}{M^2}\,,
%\end{align}
%and
%\begin{align}
%\bar{H}(M,M,M) &={}\frac{-\mu_R^2}{4 M^2} \Big(\ln^2\big(\frac{M^2}{\mu_R^2}\big) - 4 \ln\big(\frac{M^2}{\mu_R^2}\big) + \frac{5 \pi^2}{3} - 8\Big)\\
%\bar{H}(0,M,M) &={}\frac{-\mu_R^2}{2M^2} \Big(\ln\big(\frac{M^2}{\mu_R^2}\big)-\frac{\pi^2}{3}+1\Big)\\
%\bar{H}(M,0,0) &={}\frac{\mu_R^2}{4M^2} \Big(3\,\Phi(1) - \pi^2 + 2\Big)\,.
%\end{align}
%
\begin{multline}
\bar{F}_u(0,M)-\bar{F}_u(M,M) = 
\frac{3\mu_R^2}{4 M^2}\Big[\ln\Big(\frac{M^2}{\mu_R^2}\Big)-\frac{5\pi^2}{27}+\frac{1}{3}\Big]\\
%
%\shoveleft{\bar{F}_d(M,0)-\bar{F}_d(M,M) = }\\
%\shoveright{\frac{3\mu_R^2}{4 M^2}\Big[\ln\Big(\frac{M^2}{\mu_R^2}\Big)-\frac{7\pi^2}{27}+1\Big]}\\
%
\shoveleft{\bar{G}(0,M)-\bar{G}(M,M) = }\\
\shoveright{\frac{-\mu_R^2}{4 M^2}\Big[\ln^2\Big(\frac{M^2}{\mu_R^2}\Big) - \ln\Big(\frac{M^2}{\mu_R^2}\Big)+\frac{2\pi^2}{3}-\frac{9}{2}\Big]}\\
\shoveleft{\bar{H}(0,M,M)-\bar{H}(M,M,M) =} \\
\frac{\mu_R^2}{4 M^2}\Big[\ln^2\Big(\frac{M^2}{\mu_R^2}\Big) - 6 \ln\Big(\frac{M^2}{\mu_R^2}\Big)+\frac{7\pi^2}{3}-10\Big]\,,
\end{multline}
where we have suppressed the initial $\mu_{L,R}^2$ arguments of the loop functions.  The biggest contribution comes from taking the up and charm quark eigenvalues in $\tilde{Y}_u^{\dag-1}$, top quark eigenvalue in $\tilde{Y}_u^\dag$, and bottom quark eigenvalues in $\tilde{Y}_d$.  Dropping the smaller $g'$ and $\lambda_p$ contributions in \eqref{eq:HermImTrYu}, we obtain
\begin{multline}\label{eq:HermImTrYuScaling}
\argdet(\mtxY_u) \approx \frac{m_t m_b^2 m_H^2}{512 \pi^4 v^5}\sqrt{\frac{v'}{M}}\times\\
\text{Im}\Big(\sqrt{\frac{v}{m_u}}V_{ub}^* V_{tb} U_{t3}^* U_{u3} +\sqrt{\frac{v}{m_c}}V_{cb}^* V_{tb} U_{t3}^* U_{c3}\Big)\times\\
\Big[\ln^2\Big(\frac{M^2}{\mu_R^2}\Big) - 6 \ln\Big(\frac{M^2}{\mu_R^2}\Big)+\frac{7\pi^2}{3}-10\Big]\,.
\end{multline}
Taking $v'/M = 0.02$ yields
\begin{gather}
\begin{aligned}
\argdet(\mtxY_u) \approx{} & + \num{9.9e-9}\,\text{Im}[U_{t3}^* U_{u3}] \\
&+\num{4.4e-9}\,\text{Im}[U_{t3}^* U_{c3}]\,,
\end{aligned}
\end{gather}
dropping only by a factor of 2 as $v'/M$ is increased to $0.2$ (it decreases rather than increases because of the strong effects of the $\ln^2$ term).  Our already incomplete calculation in the symmetric phase breaks down if $v'/M$ is further increased, and an exact evaluation of \eqref{eq:argdetExp} in the broken phase is essential to obtain an accurate result. However,  the estimate above is much smaller than we found in the case of nonhermitian isosinglet masses, it seems reasonable to expect that in this restricted parameter space the corrections to $\bar\theta$ can be below $10^{-10}$.

\section{EDM phenomenology} 
We have argued above that the softly broken parity model does not generically solve the strong $CP$ problem because of large threshold corrections to $\bar\theta$, but it might survive in some parts of parameter space where mixing angles are relatively small or the soft parity violation is of a  special type. If we assume this is the case, it is still unlikely that $\bar\theta$ is much smaller than the present bound of $10^{-10}$, since this requires parameters to be even more restricted. The model then predicts hadronic EDMs dominated by $\bar \theta$ of this order, which is a rather specific pattern~\cite{Dekens:2014jka}. This is a generic prediction of UV solutions to strong $CP$~\cite{deVries:2018mgf}, so it  cannot be used to say anything about the mechanism of the UV-solution; moreover, it must be stressed that the same pattern of EDMs  is consistent with the SM and a small, but nonzero, $\bar \theta$ term. Nonetheless, it is an interesting and rich pattern, and these models provide motivation for experimental searches.

The archetypal observable that probes the $\bar \theta$ term is the neutron EDM $d_n$. It has been a long-standing problem to accurately calculate it as function of $\bar \theta$.  Chiral perturbation theory ($\chi$PT) can be used to calculate pion-loop contributions but the associated loops are divergent and a short-distance contributions must be included at leading order. As such, non-perturbative methods are required. Much effort has recently gone into lattice-QCD calculations of nucleon EDMs \cite{Guo:2015tla,Abramczyk:2017oxr,Dragos:2019oxn,Alexandrou:2020mds,Bhattacharya:2021lol}. The most accurate results are reported in Ref.~\cite{Dragos:2019oxn}, which performed calculations for several pion masses significantly above the physical point and several lattice spacings. A $\chi$PT fit to the pion mass and lattice spacing then gives
\begin{eqnarray}\label{Latticedn}
d_n &=& -(1.5 \pm 0.7)\times 10^{-3} \,\bar \theta\,e\, \mathrm{fm}\,.
%d_p &=& \phantom{-}(1.1 \pm 1.0)\times 10^{-3}\, \bar \theta\,e\, \mathrm{fm}\,,
\end{eqnarray}
Other lattice calculations have not confirmed these values \cite{Alexandrou:2020mds,Bhattacharya:2021lol}, and instead find results consistent with zero.  Equation \eqref{Latticedn} is consistent with QCD sum rules \cite{Pospelov:1999ha}. The proton EDM has been calculated on the lattice as well but suffered from a larger uncertainty. We will use the relation 
\begin{equation}
d_p = -(1\pm 0.5) d_n\,,
\end{equation}
which covers all existing estimates. 

EDMs of nuclei and atoms also obtain contributions from the $CP$-violating nuclear force. These are expected to be dominated by one-pion-exchange diagrams involving $CP$-odd pion-nucleon vertices \cite{deVries:2012ab} 
\begin{equation}
\mathcal L_{\pi N} = \bar g_0\,\bar N \vec \tau \cdot \vec \pi N\, + \bar g_1\,\bar N \pi_3 N\,,
\end{equation} 
in terms of the nucleon isodoublet $N = (p,n)^T$ and pion isotriplet $\vec \pi$. Contributions from heavier mesons are expected to be smaller by the $\chi$PT power counting which is based on naive dimensional analysis (NDA). NDA might break down in certain nucleon-nucleon partial waves due to fine-tuning in the scattering lengths which could indicate the need to include $CP$-odd short-distance nucleon-nucleon forces but their role is not well understood \cite{deVries:2020loy}. The pion-nucleon coupling constants $\bar g_{0,1}$ are relatively well understood \cite{deVries:2015una}
\begin{eqnarray}\label{g0theta}
\bar g_0 &=& -(14.7\pm2.3)\times 10^{-3}\,\bar \theta\,,\nonumber\\
 \bar g_1 &=&\phantom{-} (3.4\pm2.4)\times 10^{-3}\,\bar \theta\,,
\end{eqnarray}
where the smallness of $\bar g_1/\bar g_0$ is explained by the need to include additional isospin breaking from the quark masses to induce $\bar g_1$ from the $\bar \theta$ term. EDMs of nuclei and atoms can then be expressed in terms of the nucleon EDMs and $\bar g_{0,1}$. For light nuclei, this is relatively well under control by using chiral EFT for both the $CP$-conserving and -violating nuclear forces, see Refs.~\cite{Epelbaum:2008ga, Hammer:2019poc, deVries:2020iea} for reviews, and gives for ${}^2$H and ${}^3$He nuclei \cite{Bsaisou:2014zwa,Froese:2021civ}
\begin{eqnarray} \label{eq:h2edm} 
d_{{}^2\text{H}} &=&
(0.94\pm0.01)(d_n + d_p) + \bigl [ (0.18 \pm 0.02) \,\bar g_1\bigr] \,e \,{\rm fm} \, ,\nonumber\\
d_{{}^3\text{He}} &=& (0.90\pm0.01)d_n -(0.03 \pm0.01) d_p \nonumber\\
				&&+ \left[ (0.11\pm0.01) \bar g_0 + (0.14\pm0.02) \bar g_1\right] \,e  \,{\rm fm} \, .
\end{eqnarray}
Inserting the values of $d_{n,p}$ and $\bar g_{0,1}$ then gives 
\begin{eqnarray} \label{eq:h2edm} 
d_{{}^2\text{H}} &=& \phantom{-}(0.6\pm1.4)\times 10^{-3} \,\bar \theta\,e\, \mathrm{fm}\, ,\nonumber\\
d_{{}^3\text{He}}& =& -(2.5\pm1.2)\times 10^{-3} \,\bar \theta\,e\, \mathrm{fm}\,.
\end{eqnarray}
The uncertainties of these expressions are dominated by the uncertainties in the QCD matrix elements linking $\bar \theta$ to $d_{n,p}$ and $\bar g_1$ which will be improved in future lattice calculations. EDMs of diamagnetic atoms are, right now, more relevant but suffer from nuclear uncertainties. For instance \cite{Engel:2013lsa,Yamanaka:2017mef,Dobaczewski:2018nim}
\begin{eqnarray}\label{dHg}
d_{\rm Hg}&=& -(1.8\times 10^{-4})\bigg[(1.9\pm0.1)d_n +(0.20\pm 0.06)d_p\nonumber\\
&&\times \bigg(0.13^{+0.5}_{-0.07}\,\bar g_0 + 0.25^{+0.89}_{-0.63}\,\bar g_1\bigg)e\, {\rm fm}\bigg]\,,\\
d_{\mathrm{Ra}} &=& (7.7\times 10^{-4})\left[(2.5\pm 7.5)\,\bar g_0 - (65 \pm 40)\,\bar g_1\right]e\, {\rm fm}\,,\nonumber
\end{eqnarray}
where the nucleon EDM contributions to ${}^{225}$Ra have been neglected. The overal factors in front of these expressions are atomic screening factors that are relatively well under control. For $\bar \theta$-dominated EDMs we then obtain 
\begin{eqnarray} \label{eq:h2edm} 
d_{\rm Hg} &=& \phantom{-}(0.4\pm1.2)\times 10^{-6} \,\bar \theta\,e\, \mathrm{fm}\, ,\nonumber\\
d_{\rm Ra} &=& -(2.8\pm2.8)\times 10^{-4} \,\bar \theta\,e\, \mathrm{fm}\, .
\end{eqnarray}
These expressions have large uncertainty due to the nuclear matrix elements but still provide good tests for a $\bar \theta$-dominated scenario. Most higher-dimensional sources of $CP$ violation would lead to rather different ratios of EDMs. 

It was recently pointed out that the dramatic increase in the sensitivity of EDM measurements of polar molecules also has impact on the $\bar \theta$ term \cite{Flambaum:2019ejc}. The idea is that a nonzero $\bar \theta$ term leads, through electromagnetic loops, to $CP$-odd electron-nucleon interactions that, in turn, induce molecular EDMs. Ignoring the contribution from the electron EDM, which is very small in $\bar\theta$-dominated scenarios, the quantity of interest are frequencies \cite{doi:10.1063/1.4968597,Skripnikov,PhysRevA.96.040502,doi:10.1063/1.4968229,PhysRevA.90.022501,PhysRevA.93.042507}
\begin{eqnarray}
\omega_{\text{YbF}} &=& -(17.6\pm2.0)(\mathrm{mrad}/\mathrm{s})\left(\frac{C_S }{10^{-7}}\right)\,,\nonumber\\
\omega_{\text{HfF}} &=&\phantom{-}(32.0\pm1.3)(\mathrm{mrad}/\mathrm{s})\left(\frac{C_S }{10^{-7}}\right)\,,\nonumber\\
\omega_{\text{ThO}} &=&\phantom{-}(181.6\pm7.3)(\mathrm{mrad}/\mathrm{s})\left(\frac{C_S }{10^{-7}}\right)\, ,
\end{eqnarray}
associated to reaction of the molecule due to external electric fields. 
where $C_S = C_S^{(0)}+\frac{Z-N}{Z+N} C_S^{(1)}$ in terms of $Z$ and $N$ of the heaviest atom of the molecule, and the electron-nucleon interactions
\begin{equation}\label{eN} 
\mathcal L =-\frac{G_F}{\sqrt{2}}\,\bar e i\gamma_5 e\, \bar N\left(C_S^{(0)}+\tau_3 C_S^{(1)}\right) N\,.
\end{equation}
The $Z$-to-$N$ ratio is rather similar for these systems and Ref.~\cite{Flambaum:2019ejc} effectively finds $C_S = -(3\pm1.5)\times 10^{-2}\,\bar \theta$. The most stringent limit is $\omega_{\text{ThO}} <1.3$ mrad/s \cite{Andreev:2018ayy}, then leads to $\bar \theta < 3 \times 10^{-8}$ which is not yet competitive with the neutron EDM limit but also not very far away.

\section{Summary}
We have revised the estimate of two-loop corrections to $\bar\theta$ in the minimal parity-based solution to the strong $CP$ problem. We find larger corrections than previously accounted for, and no suppression by the weak scale. Soft parity breaking in the mass matrices for the isosinglet quarks leads to contributions to $\bar\theta$ of order $10^{-3}$ times mixing angles and phases. Soft parity breaking in the Higgs sector leads to much smaller corrections which we estimate to be of order $10^{-8}$ times mixing angles and phases. If the parameters can be restricted so that present bounds are evaded, $\bar\theta$ is still not likely to be far below $10^{-10}$, so the model presents an interesting target for future EDM experiments.

\section*{Acknowledgments}
We thank Isabel Garcia-Garcia for useful discussions. PD and HHP acknowledge support from the US Department of Energy under Grant numbers DE-SC0015655 and DE-SC0010107, respectively.
\bibliography{references}{}

\end{document}